\documentclass[10pt,letterpaper]{article}
\usepackage[top=0.85in,left=2.75in,footskip=0.75in]{geometry}

\usepackage{amsmath,amssymb}

\usepackage{changepage}

\usepackage{textcomp,marvosym}

\usepackage{cite}

\usepackage{nameref}
\usepackage[hidelinks]{hyperref}

\usepackage[right]{lineno}

\usepackage[nopatch=eqnum]{microtype}
\DisableLigatures[f]{encoding = *, family = * }

\usepackage[table]{xcolor}

\usepackage{array}

\newcolumntype{+}{!{\vrule width 2pt}}

\newlength\savedwidth

\newcommand\thickhline{\noalign{\global\savedwidth\arrayrulewidth\global\arrayrulewidth 2pt}%
\hline
\noalign{\global\arrayrulewidth\savedwidth}}

\raggedright
\usepackage[aboveskip=1pt,labelfont=bf,labelsep=period,justification=raggedright,singlelinecheck=off]{caption}

\makeatletter
\renewcommand{\@biblabel}[1]{\quad#1.}
\makeatother

\usepackage{lastpage,fancyhdr,graphicx}
\usepackage{epstopdf}
\fancyheadoffset[L]{2.25in}
\fancyfootoffset[L]{2.25in}
\begin{document}
\vspace*{0.2in}

\begin{flushleft}
{\Large
\textbf\newline{Emerging STEM education researchers’ positioning and perception of discipline-based education research} 
}
\newline
\\
Shams El-Adawy\textsuperscript{1},
Cydney Alexis\textsuperscript{2},
Eleanor C. Sayre\textsuperscript{1,3}
\\
\bigskip
\textbf{1} Department of Physics, Kansas State University, Manhattan, KS, USA
\\
\textbf{2} Department of English, Kansas State University, Manhattan, KS, USA
\\
\textbf{3} Center for Advancing Scholarship to Transform Learning, Rochester Institute of Technology, Rochester, NY, USA
\\
\bigskip

%
%






\end{flushleft}
\section*{Abstract}
Various motivations bring researchers to discipline-based education research (DBER), but there is little research on their conceptualization of and navigation into this new-to-them area of research. We use phenomenography to analyze interview data collected from twenty-eight emerging STEM education researchers to gain a better understanding of how they perceive themselves within DBER and what they perceive it to be.  Grounded in the figured worlds theoretical framework, we identify the spectrum of ways emerging STEM education researchers identify or project themselves into this new space: to improve their teaching, to make it their new primary research field, and/or to negotiate how it will fit with their primary one. We also highlight salient negotiations  that emerge because of the close ties between DBER and disciplinary science, which provides us with a better understanding of emerging researchers' perceptions. This work generates insight into the kinds of professional development opportunities that would support emerging education researchers within STEM departments and the broader DBER community.



\section*{Introduction}
Discipline-based education research (DBER) is an interdisciplinary field that investigates discipline-specific learning and teaching, often paired with more general research on human learning and cognition \cite{national_research_council_us_discipline-based_2012}. DBER grew from the individual STEM disciplines’ interest in investigating how to improve learning and teaching. To better understand how discipline-based research  developed, we are interested in the professional trajectories of those who do it. In particular, we want to understand how classically-trained scientists in university faculty positions embark on research in this area. 

Studies of STEM faculty who engage in DBER have generally focused on Science Faculty with Education Specialities (SFES) \cite{bush_investigation_2011, bush_evolving_2019}. SFES have been conceptualized as individuals contributing to STEM education reform from a wide variety of academic position types in STEM departments \cite{bush_disciplinary_2020}. Most SFES work examines the experiences of faculty-level researchers hired to do education research. However, the SFES category does not count all people who do DBER. For example, classically trained STEM faculty at small institutions can start doing it  as part of their institutions’ pedagogical missions. Their experiences are not captured within the SFES category because they are neither hired to do education work nor is their work in education research a central scholarly feature of their jobs. In parallel, although there is an increase in DBER graduate degrees and postdocs, a substantial portion of researchers do it without formal training (i.e. with neither graduate degrees nor postdoctoral positions in DBER). These researchers are what we call ``emerging STEM education researchers,” the population of interest in this paper. This group of emerging STEM education researchers need support as they enter the field.          

Furthermore, individual STEM disciplines have investigated professional development in their communities. Researchers in mathematics, biology, and physics education have (separately) examined the impact of interventions and programs on the professional development of their faculty \cite{archie_investigating_2022,corbo_framework_2016, dancy_faculty_2019, chasteen_insights_2020}. Given the various communities faculty engage in and experiences they bring, these discipline-specific conceptualizations focus on particular facets of their professional development. The mathematics education research community has investigated the different ways communities of practices can support faculty in making changes to their teaching \cite{johnson_supporting_2021}. The biology education research community has studied the tensions in scientists’ professional identity and teachers’ pedagogical reform \cite{brownell_barriers_2012}. The physics education research community has emphasized the importance of an agentic and holistic approach to professional development of physics faculty \cite{el-adawy_context_2022, strubbe_beyond_2020}. Across these different studies, the focus has been on how to support faculty in improving their teaching, rather than on supporting them in engaging in DBER research.  This focus is mirrored by the larger field of faculty professional development, which centers questions of improving faculty teaching over research on improving or supporting faculty scholarship.
 
More research is needed on how these emerging STEM education researchers enter the field and the kinds of support they need. Thus, we center our research study on participants who are interested in doing DBER, whether or not they were hired to do so, and we focus on their professional development as researchers in this space. In particular, we are interested in better understanding how emerging STEM education researchers project themselves within this community and how their professional identities evolve as a result of their engagement.

The spectrum of reasons that bring researchers to DBER is valuable to investigate as it allows us to better understand their experiences as emerging STEM education researchers, their interactions with this new community, and the diverse ways they perceive its role as a research area. We take up an identity and positioning frame to better understand the adjustment of these emerging scholars. We address the following research question: \textbf{How do emerging STEM education researchers position themselves and perceive the role of discipline-based education research?}

To model how emerging STEM education researchers conceptualize this transition, we turn to Figured Worlds, an identity and positioning theory.  Figured Worlds models how researchers' perceptions of professional growth interact with perceptions of their environments.  We interview twenty-eight emerging STEM education researchers. Our analysis identifies three dominant ways they envision this research field in their professional lives: to improve their teaching, to make it their new primary research field, and to negotiate how this new research field will fit with their primary one. In turn, these results suggest insights for supporting this community’s professional development.

\section*{Theoretical framework: Figured Worlds}
To explore how emerging STEM education researchers position themselves and perceive the role of discipline-based education research, we draw upon an identity and position framework, figured worlds. This framework centers individuals’ interpretation of their experiences as a result of their community involvement. Figured worlds is a theory that captures how individuals imagine or identify their identity and position within a particular social, cultural, and historical context \cite{holland_identity_1998}. This framework is compelling  because it sees identity formation as continually evolving, due to interactions within an individual's various sociocultural spaces. The identity formed within a figured world comes from participation in its activities and from processing the meaning of one’s identity in shifting and evolving socio-cultural contexts.   
The framework underlines four features that characterize the figured world of an individual or a group of individuals as they negotiate their identity \cite{urrieta_figured_2007}. Recruitment and development refer to how learners process entry and growth in a new sociocultural space. Meaning creation is about how they make sense of the space’s norms. Positioning is about how they situate themselves and their potential contributions. Social organization is how they perceive the power dynamics at play. Table \ref{tablefw} summarizes the components of the framework and they way we contextualized them for this study. 

\begin{table}[!ht]
\centering
\caption{
{\bf Components of figured worlds: the theoretical and contextualized definitions.}}
\begin{tabular}{|m{2.9cm}|m{4.2cm}|m{4.5cm}|}
\hline
\textbf{Component of Figured World} & \textbf{Framework definition }  & \textbf{Contextualized definition} \\ \thickhline Recruitment and development  & How learners process entry and growth in sociocultural space & How emerging STEM education researchers are recruited in DBER and how they engage with DBER \\ \hline 
 Meaning creation & How learners make sense of the sociocultural space’s norms & How emerging STEM education researchers  make sense of practices and activities in DBER  \\ \hline
 Positioning  & How learners situate themselves and their potential contribution in socio-cultural space & How emerging STEM education researchers identify themselves within DBER \\ \hline Social organization & How learners perceive the power dynamics at play  in socio-cultural space & How emerging STEM education researchers conceptualize relationships, expectations and values in DBER
\\ \hline
\end{tabular}
\label{tablefw}
\end{table}

Individuals are often part of multiple figured worlds that come together to shape experience. There is a range of ways that the figured world can be articulated: groups of professionals, particular classrooms, and particular institutions can all be characterized as figured worlds. For example, universities are spaces grounded in discourses and practices that are socially, culturally, and historically shaped and in which academic and disciplinary identities are formed \cite{zuckerman_transfer_2021}. 

In her work, Holland emphasizes that identity formation occurs in practice and is formed and reformed through activities and events that individuals part take in \cite{holland_identity_1998}. She does not centralize demographic identity such as gender and  race. Instead, she focuses on the development of identity in relation to the practices and activities individuals engage in. Holland states that of significance to the concept of figured worlds “is the situatedness of identity in collectively formed activities. These identities that concern us are ones that trace our participation, especially our agency, in socially produced culturally constructed activities”\cite{holland_identity_1998}. 

Holland distinguishes between three types of identities: relational, positional, and figurative. The interplay of these three types shape and reshape the figured world of an individual. Relational identity, often mediated through speech/communication, refers to who we are in relation to our interactions with others. Positional identity refers to one’s apprehension of their social position relative to others and socio-cultural structures. Figurative, or narrativized, identity is about one's perception of themselves and others in a given cultural world. The imaginative and identification aspects of the figurative lens are some of the features of identity that are helpful to this project and focus our analysis on individuals' processing and perceptions as they conceptualize their navigation into new fields. 

STEM emerging education researchers develop figured worlds based on their interactions with the DBER field in terms of the four characteristics of the framework (recruitment and development, meaning creation, positioning, and social organization). Individuals are recruited or enter this field and it evolves as a result of its members’ work.  DBER practices and activities create meaning for members. The field is socially organized; people learn how to relate to each other within the space based on what is expected and valued. Lastly, individuals identify themselves within the space based on actions taken, as well as the field’s discourse.

When individuals explore new research areas such as DBER, they see their identities as in flux and they see potential for how they might grow as teachers and scholars. Their identities are constructed within dynamic social and cultural worlds, including academic networks within their  home discipline and the broader research community. Thus, when exploring new research areas, discussions about one’s professional identity are at play, including the interplay of relational, positional, and figurative identities. Professional identities are constructed within the social and cultural world of academia, and often multiple professional identities are intertwined and impact one’s professional development \cite{jiang_understanding_2021}. Within the broader world of academia,  a person’s sense of disciplinary identity encompasses how they  understand themselves, interpret experiences, present themselves, wish to be perceived, and are recognized by the broader professional community \cite{kane_young_2012}. Using figured worlds allow us to focus, then, on how individuals harness and process their professional identity development as they begin to engage in DBER. 

To sum up, the value of this framework is its illustration of the new and different possibilities that individuals figure for themselves within the worlds they are part of \cite{calabrese_barton_crafting_2013}. The use of figured worlds allows us to articulate emerging STEM education researchers' conceptualization of and navigation into DBER. Particularly, it looks at how their entry is shaped by their own histories, their involvement in professional development opportunities, and the larger context of their social and cultural environment. By examining what brings researchers to DBER, particularly their figured world, we illustrate what is perceived as important to newcomers as they negotiate entry into a new research field. In turn, it provides the community with knowledge of how newcomers perceive the field. It also provides the opportunity to address newcomers’ perceptions and challenges.

\section*{Context}
Our study includes emerging STEM education researchers who participated in a professional development program, named Professional development for Emerging Education Researchers (PEER)\cite{el2023professional}. PEER is a professional development program designed to help faculty, postdocs, and graduate students jumpstart their transition into the world of discipline-based education research. The central activity of the PEER program is intense experiential workshops to help emerging STEM education researchers develop quality research projects; engage in targeted experiential work to develop their projects and skills; and collaborate and form a support community of peers, mentors, and collaborators. Our dataset also included emerging STEM education researchers that were not involved in PEER but were transitioning into education research. They were recruited from our engagement with professionals expressing interest in making this transition
 
This study tracked twenty-eight emerging discipline-based education researchers who had  a disciplinary emphasis in mathematics, physics, and biology. Participants were solicited for semi-structured interviews, during which  we asked about their DBER experiences. Interviews for all research participants were conducted and recorded by the first author and another member of the research team over video conference (Zoom), and then a professional transcriptionist transcribed the interviews .

\section*{Methodology}
We analyzed our interview data using a phenomenographic approach. Phenomenography is a research methodology used to examine how individuals experience a phenomenon \cite{marton_phenomenographyresearch_1986}. Developed within educational environments, its goal  is to describe the variation in people’s experiences around any given phenomenon. It has been used in physics education to investigate students’ learning experiences in the physics classroom \cite{williams2018physics} and their identity development in the field \cite{irving2013physics, irving2015becoming}. Phenomenography has also been used to characterize physics faculty’s beliefs and approaches to instructional change  \cite{alaee2020processes, huynh_building_2021}. More narrowly, researchers have  examined students’ problem solving approaches in introductory physics\cite{walsh_phenomenographic_2007} and to characterize students’ conceptual understanding of particular physics topics such as electric and magnetic interactions \cite{hernandez_phenomenographic_2022}.  Outside the DBER context, phenomenography has been used in higher education to study how academics conceptualize research, particularly capturing the different ways faculty and graduate students understand the nature of academic work \cite{brew_conceptions_2001, akerlind_academic_2008}.

This methodology allows us to surface ideas that individuals have about the phenomenon studied  \cite{hajar_theoretical_2021}. In doing so, we gain insight into their experiences. In turn, it can uncover contradictions in reasoning and open up the possibility to consider alternative ideas\cite{marton_phenomenographyresearch_1986}. Additionally, phenomenography can be useful when examining a phenomenon that is hard to define, complex, or could have a variety of meanings, \cite{cossham_evaluation_2018} because it can highlight the diverse ways people experience a particular transition or learning experience. Raising awareness of this variation opens the possibility to be more inclusive.    

The phenomenon we are exploring is the transition into DBER at different career stages and in different contexts. We use phenomenography to capture the myriad ways emerging STEM education researchers conceptualize this field and their role in it. Pairing phenomenography with the figured worlds theoretical framework allows us to uncover typical stories around DBER’s socio-cultural context that  “are usually unconscious and taken-for-granted”\cite{gee_how_2011}. This is an important strategy to gain a better understanding of the experiences of emerging STEM education researchers and to identify opportunities for more inclusive and supportive professional development.

During our initial data analysis, we repeatedly read the interviews and identified categories of experiences across interview participants. We  grounded our search for categories in the components of the figured worlds framework (recruitment and development, meaning creation, social organization, and positioning). We looked for how participants talked about their experiences as it relates to these components. As we developed categories, our purpose was to see what qualitatively different categories of experience emerged as significant. We noted that some participants had a diversity of experiences, which meant that some individuals could fall in qualitatively different categories of experiences. 

After category identification, we created codes and sought nuances within categories to identify subcategories. Moving back and forth between data and categories of experiences led to the creation of the initial set of codes and definitions that formed the codebook. The first author gathered all the quotes and characterized them in the codebook, refining definitions and codes continuously. The first author provided the codebook and 10\% of the quotes to another member of the research project’s team to check the reliability of the codes, definitions and characterization of quotes. This IRR researcher independently coded these quotes. Before discussion, there was 72\% agreement between the first author and the IRR researcher. Discussion occurred, which consisted of providing more context from data about the quotes and refining the language and meaning of the codes. Afterwards, the first author and IRR researcher reached 98\% agreement.  Lastly, the analysis was collaboratively discussed until consensus developed among the project’s researchers and extended research team.

\section*{Analysis}
To answer our research question: \textbf{How do emerging STEM education researchers position themselves and perceive the role of discipline-based education research?} we draw upon all four features of the figured worlds framework. To characterize which categories participants fell into, we focused on participants’ self-identification. The study  identified three ways they see their roles in education research: to improve teaching (Improvers), to join a new field of research (Joiners) or negotiate their position and identity in DBER vis-à-vis their home discipline (Negotiators).

For Improvers, the DBER figured world is about improving teaching by engaging in education research. For Joiners, the DBER figured world is conceptualized as a space they want to fully engage and grow in as professionals and not just to fulfill current job responsibilities. Negotiators conceptualize DBER as a space for exploration. However, they are still figuring out how to navigate the cultural and procedural norms between their home discipline and DBER. We identified across our dataset these three qualitatively different experiences, yet some of our participants identified  with multiple categories. In other words, categories are not mutually exclusive. For the purposes of this study, in which we want to delineate initial categories as clearly as possible, we placed individuals into the category that best fit their dominant perceptions about this new research area. All of the participants referenced in the following sections are unique study participants, unless noted otherwise.

\subsection*{Improvers}
Participants in the Improvers category see the figured world of DBER as an opportunity to improve practice locally in an evidence-based way. Within the improvers category, we found three sub-categories, those who primarily wanted to improve classroom practices, those who wanted to improve department practices and those wanting to improve service work.

\subsubsection*{Classroom practice improvers}

This subset of improvers  wants to enhance STEM classroom practices to increase student learning. For example, one participant, an assistant professor of mathematics at a U.S. private institution, shares what drew them to DBER:
\begin{quote}
    \textit{desire to better serve my students leads me to want to analyze in some more rigorous way how I’m serving them and how to better do so and use evidence-based strategies}
\end{quote} 
First-hand classroom experiences  often lead to interest in education research. Nevertheless, Improvers in this category highlight the opportunity to act upon their classroom observations  in a scholarly way. The meaning creation they are assigning to their DBER figured world stems from assigning significance and importance to improving teaching. As such, they see this new field  as an opportunity to simultaneously take up the roles of teacher, researcher, and participant. DBER allows them to be more intentional and thoughtful in their decision-making processes.

Although they want to approach instructional change with a research-based approach, their lack of familiarity with established DBER researchers may translate into low self-efficacy. For example, an associate professor of mathematics  at a public U.S. research university shares some of their struggles:
\begin{quote}
    \textit{I feel like a collaborator who really knows what they’re doing would be helpful. I feel like as I’m doing this, I’m kind of just making it up as I go.}
\end{quote}
As a new member, this need for collaborators, mentors and peers on education projects speaks to the community support emerging STEM education research need \cite{hass2021community}. Particularly, this past research shows that these support roles can help build and maintain emerging STEM education researchers' confidence, thus promoting their growth within the community.

\subsubsection*{Departmental teaching practice improver}

This type of improver wants to make research-based changes in their department. One participant, who is a full professor of mathematics at a public U.S. university says:
\begin{quote}
\textit{I want our department, and myself as a teacher, to make good choices about what we do in the classroom, how we structure our curriculum, that sort of thing. So I’m interested in making informed decisions and making those decisions in a context in which we can decide if those are good decisions or not.} 
\end{quote}
The scope of this emerging STEM education researcher's DBER figured world is broader than just their classroom environment, it is department-wide. However, their figured world remains a means by which to improve their teaching. Despite this department's goal to improve teaching,  another emerging STEM education researcher, a full professor of mathematics at a U.S. public university, describes the lack of formal training in education research as a perceived barrier to transition into the community : 
\begin{quote}
\textit{I think I have contributions to share with the RUME [Research in Undergraduate Mathematics Education] or the PERC [Physics Education Research Conference] community, and I’m not trained in that. And so sort of being able to do education research or talk with people that do education research seems like a requirement to share those ideas with those communities.}
\end{quote}
This challenge is described as a lack of familiarity with norms and resources, but also underlines the image the DBER community projects to emerging researchers about recruitment and development. Extensive formal training is seen as necessary  before engaging with DBER at any level. 

\subsubsection*{Student-centered service improvers}

This improver wants to do DBER  to enhance the impact of their service work, which they foresee as translating to their classroom environment. A participant, who is an associate professor of mathematics at U.S. public university is interested in looking at the 
\begin{quote}
    \textit{impact that outreach may have on students’ identity as well as students’ ability to be successful in their mathematical work within the classroom. So trying to make that connection to how that enrichment experience impacts their long-term academic success.}
\end{quote} 
The service part of their faculty role is central in informing their interest in DBER. They imagine DBER as a tool to help them enhance the effectiveness of informal learning environments on students’ science identity. As a by-product, this can help improve their sense of belonging in STEM  classrooms. However, this emerging STEM education researcher, the same associate professor of mathematics cited the  classroom improvers subcategory, faces resistance as they engage as an emerging STEM education researcher in their department: 
\begin{quote}
    \textit{My chair told me earlier this summer that, you know, `education research should be done in the college of education. That’s what they’re for, right? We’re a department of mathematics and statistics, so we should do mathematics and statistics.'}
\end{quote}
DBER is not valued or supported in their STEM department; it is not viewed as the same caliber as their research in their primary discipline. As a result, the social organization of their figured world contains strife with members outside of DBER, which cannot help but shape their perception of it.  Despite the meaning and value they see in education-based research,  the resistance they face from  their home department challenges their transition.

Although there are nuances in Improvers’ conceptualization of their figured worlds, their DBER world  is conceptualized as space that stems from their professional work as teaching faculty and department members. However,  their imagination and identification are challenged by existing interactions with their departments.  These tensions are shaping their transition. This category of experiences illustrates the need for discussions and support structures that center around the intellectual and personal benefits from divergences in and broadening the scope of   professional trajectories.

\subsection*{Joiners}

In contrast to Improvers who conceptualize DBER as a space that stems from their professional work as teaching faculty and department members, Joiners conceptualize it  as a space they want to fully engage in and grow in as professionals. They do not view it just as a means to improve teaching practices at their institutions; instead,  they want to fully engage in a community  beyond their department and institution. Their transition process and development is facilitated by their interactions with the community and their institutions’ expectations. Most participants with Joiner characterization found themselves exclusively in this category, yet some participants had Improvers characteristics as well. For the purposes of this study, as indicated above, we focused on aligning participants with the category they predominantly identified with, unless otherwise noted.

\subsubsection*{Joiners seeking interdisciplinary connections}

This emerging STEM education researcher joins  because they want to fully engage with this research area. They have had encouraging interactions with the community, which positively fuels their recruitment and development. One participant, who is an associate professor of physics at a U.S. undergraduate institution says:
\begin{quote}
    \textit{The  physics education research field has been so welcoming to just someone who’s an outsider that’s just curious. It seems like a welcoming field and that’s one thing that’s drawn me to it and my burgeoning interest in doing this research myself.}
\end{quote} 
The positive interactions with the community opened up the possibility for them to identify with the field. The processing of their recruitment and development into this figured world helped  them to imagine themselves part of the community. Similarly to Improvers, Joiners have a robust conceptualization of what meaning they assign to DBER engagement. For example, another emerging STEM education researcher in this subcategory expands on this reasoning. They articulate the value they see in connecting multiple disciplinary identities together to offer insight into the world they are joining: 
\begin{quote}
    \textit{I’m really interested in, you know, what I am bringing as a mathematician to the understanding of math education research.}
\end{quote} 
They are processing their figured world by imagining the different possibilities of connecting ideas, concepts, and perspectives from traditional STEM disciplines and education research. The combination of encouraging DBER interactions and the interdisciplinary nature of the work shape their emergent  figured world. On the one hand, they are enjoying the playful and creative ways of knowing and understanding  that can come from interdisciplinary work. On the other hand, they are encouraged by positive community interactions.

\subsubsection*{Joiners seeking action-oriented research}

Supported by their institution's teaching focus, this emerging STEM education researcher imagines integrating discipline-based research into their research portfolio. For instance, an associate professor of mathematics at U.S. public university within this category describes how their institution is providing space to engage in DBER:
\begin{quote}
    \textit{I’m at a teaching-focused institution, which means that research is very broadly defined, research can include math education research or teaching-specific research as well as undergraduate-led research, and then also the traditional kind of math research papers in pure mathematics or applied mathematics research.}
\end{quote}  
Exploration of this new research area is possible because their current institutional environment is focused on improving teaching and supports a broad category of scholarship. In turn, this gives them the flexibility to engage in new research areas. Their institution promoting and valuing a broad range of activities within the umbrella of scholarship encourages their engagement.

Another participant who is an instructional faculty in mathematics at public U.S. research university expands beyond the institution's mission and elaborates on the value of the tangible results that are possible:
 \begin{quote}
    \textit{One thing that was exciting to me about that is kind of the feeling of how, this is actually an impactful career, like this is something that’s making a difference, a very tangible difference.} 
\end{quote}
The direct application of results can benefit student learning and improve their experiences in STEM courses, which in turn gets them excited about offering meaningful contributions.

Joiners imagine DBER  to be a new primary research field. They are energized by the opportunity to explore its interdisciplinary and engage in applied research. Encouraged by support from their institution and the community, the multiple possible lenses to approach research problems creates excitement. They see value and meaning in purposefully aligning their research on instructional change to their classroom practice to improve student experiences.

Comparing Improvers and Joiners, we notice that the similarity between these two categories lies within their  robust conceptualization of DBER and especially the potential impact of their engagement. For Improvers, it is a tool to achieve their current needs as faculty in a research-based way. For Joiners, it is a new area of research that will help them  expand their research portfolio. Although many of our participants had robust conceptualization of DBER, some articulated how they were negotiating some features of each. This led to the identification of the Negotiator category which highlights ongoing change in their perception of discipline-based research.

\section*{Negotiators}

In contrast to the robust perception and role of DBER for Improvers and Joiners, the third category, Negotiator, brings to the forefront an unsettlement of where DBER fits within their professional lives. The Negotiator is grappling to find their position and identity in both fields. The Negotiator category highlights three kinds of salient negotiations emerging STEM education researchers engage in as they conceptualize entry in DBER: (1) positioning within this new field, (2) professional identity within this new field, and (3) positioning within both  DBER and  STEM departments. Participants in the Negotiator category also identified with features of the Improver and Joiner categories; because of this, quotes within this category sometimes came from the same  interview participants, and we indicate these instances below.

\subsubsection*{Negotiating positioning}

Some Negotiators want to improve their teaching and engagement with DBER. A full professor of mathematics at a public U.S. institution who was also cited in the departmental teaching practice improver category reflects on the ways to position himself. He wonders, \textit{“What’s the way to work with education research or what’s the overlap”} between doing DBER himself or collaborating with DBER researchers as an engaged instructor, and how he can be \textit{“part of that community”}. He is trying to figure out how to collaborate with education researchers in productive ways but is unsure of what expertise DBER  values and what a STEM researcher can contribute. The act of figuring this out contributes to the creation of the individual’s figured world \cite{holland_identity_1998}. However, the navigation of norms in an interdisciplinary field such as DBER is daunting to emerging scholars, even for  researchers experienced in other areas. Creating and finding productive research partnerships is how researchers come to understand how a research field organizes itself. However, this negotiator sees a challenge in navigating  collaboration, which impacts both  how they position themselves  and their conception of its  figured world.

\subsubsection*{Negotiating professional identity}

This subcategory of researchers is negotiating which aspects of their professional life drive their DBER engagement. In the interviews, one of the participants who is a mathematics instructional faculty  at a U.S. research institution who is also cited in joiners seeking action-oriented research subcategory reflects on her professional identity as having tension between mathematics and mathematics education research: 
\begin{quote}
    \textit{I was a mathematician, but no, I'm not a mathematician anymore. Oh, and maybe I'll be a math ed researcher? No, I’m not really a math ed researcher either.}
\end{quote}
This researcher no longer identifies with her past experiences, particularly her training as a mathematician, which influences her lack of identification as a mathematics education researcher. This back-and-forth regarding which disciplinary identity fits is brought to the forefront in the DBER space. While this negotiator sees these two identities as sequential and exclusive, other identity negotiators want to keep both professional identities active. DBER identity becomes a mediating force that causes them to reconsider and renegotiate their professional identity. Although it is common to belong to multiple figured worlds and for figured worlds to evolve over time \cite{zuckerman_transfer_2021}, this subcategory of negotiator stresses that juggling among practices and activities in different figured worlds leads to a challenge in their conceptualization and navigation into the DBER figured world, and it communicates important information about how the emotions and thoughts people experience as they transition into new scholarly roles and spaces.

\subsubsection*{Negotiating tension between DBER and traditional STEM disciplines}

As a field, DBER promises to improve teaching and learning \cite{national_research_council_us_discipline-based_2012}; however, many STEM faculty are skeptical that its results are as robust as decades of teaching experience \cite{madsen_research-based_2016}. This tension between research and practitioner expertise \cite{coburn_researchpractice_2016} complicates DBER figured worlds for our participants. One tenured math professor at a mainly undergraduate U.S. institution, who is also cited in the student-centered service improvers subcategory, stresses that some senior mathematicians and faculty \textit{“don’t value math education research.”} She imagines senior mathematicians saying \textit{“you can’t possibly capture what I know from my two decades, three decades, five decades of experience. Like, math education researchers just can’t do it.”} Part of her figured world of mathematics includes devaluing education research results, as embodied in these imaginary-yet-powerful experienced mathematicians. As this subcategory of negotiator imagines herself engaging more in DBER, she anticipates discovering more of \textit{“those culture fights that have been existing that you don’t even realize that you are stepping into.”}  This is a common tension that exists among interdisciplinary education research \cite{peffer_practical_2016}, but it is especially challenging to navigate for individuals getting started in DBER. 

For individuals in this category, negotiations highlight different aspects of DBER as a figured world. They want to form collaborative interdisciplinary relationships, but also worry about  how expertise is valued. In contrast, negotiating identity focuses on how individuals experience the act of taking up new research identities.  Finally, Negotiators who focus on the tension between education research and practitioner expertise are concerned that joining might obligate them to fight cultural battles they do not yet understand; their figured world includes potential strife with powerful senior faculty. 

Unlike Improvers and Joiners, Negotiators are in flux; they are grappling with making a shift in their current professional roles. Nevertheless, we notice a similarity in challenges  that Improvers and Negotiators face in contrast to Joiners. On the one hand, Improvers and Negotiators are grappling with how to position themselves in DBER as trained scientists and how to manage existing tensions between the two  disciplines. On the other hand, Joiners are encouraged by the ease of finding collaborators within the welcoming DBER community and the support of their departments and institutions for interdisciplinary engagement.

\section*{Results and Discussion}

Using the figured worlds framework brings to the surface the underlying ideas and conceptions that emerging STEM education researchers may have about DBER in terms of recruitment and development, meaning creation, social organization, and positioning. Emerging STEM education researchers conceptualize their DBER figured world with nuance, as summarized in Table \ref{tablesummary}. There is a spectrum of experiences shaping  emerging STEM education researchers’ the DBER figured worlds. Improvers, Joiners, and Negotiators discuss challenges and opportunities that are related and can be mutually reinforcing. Individual participants could appear in multiple categories, and their figured worlds grow and change as they navigate these tensions and their specific professional contexts and goals.

\begin{table}[!ht]
\centering
\caption{
{\bf Nuanced experiences of emerging STEM education researchers projections of themselves within DBER }}
\begin{tabular}{|m{2cm}|m{3.3cm}|m{3.3cm}|m{3.3cm}|}
\hline
\textbf{Category of  Experience} & \textbf{Improver}  & \textbf{Joiner} & \textbf{Negotiator} \\ \thickhline
 & Classroom practice improvers  & Joiners seeking interdisciplinary connections & Negotiating positioning \\ \hline 
  & Departmental teaching practice improver & Joiners seeking action-oriented research  & Negotiating professional identity \\ \hline
  & Student-centered service improvers & & Negotiating tension between DBER and traditional STEM disciplines
\\ \hline
\end{tabular}
\label{tablesummary}
\end{table}

\subsection*{Challenges and opportunities across categories of experiences}

Emerging STEM education researchers face challenges and opportunities as a result of their perceptions: the DBER dichotomy (a divide between supportive and discouraging interactions with DBER) and the STEM department dichotomy (a divide between supporting and discouraging DBER in STEM department). There are subtle differences and similarities across categories of experiences. Yet, one overarching pattern is that emerging STEM education researchers who identify with the Joiner category tend to find themselves in supportive STEM departments and have encouraging DBER community interactions, whereas Improvers and Negotiator have the opposite experience. 

Improvers and Negotiators face similar challenges in terms of imagining and identifying their position within the DBER figured world. They are grappling with how to position themselves as trained scientists and how to manage existing tensions between these two  disciplines. The major difference between Improvers and Negotiators lies in where they are at in their conception of their DBER figured world. Improvers imagine DBER as a tool to improve existing responsibilities and the challenges and hurdles they face in their navigation of DBER, whereas Negotiators imagine the possible tensions as they grapple with whether they want to use DBER for improvement of current responsibilities or to join a new research community. 

Nevertheless, the significant correspondence between the challenges of Improvers and Negotiators suggests that DBER is hard to engage with because of newcomers’ two challenging perceptions: the perception of legitimacy of DBER within STEM departments and gatekeeping by the broader DBER enterprise. The latter is concurrent with findings from other researchers investigating emerging STEM education researchers' sense of belonging \cite{hass_gatekeeping_2023}. As such, the DBER community needs to advocate for the value of education research within disciplinary STEM department. DBER also needs to articulate more explicitly the norms of what it takes to do DBER by providing resources and training for researchers interested in engaging in this research area. 

Comparing Improvers and Joiners, we notice that the similarity between these two categories lies within the fact they have a robust conceptualization of what their DBER figured world might be. Nevertheless, a notable distinction between Joiners and Improvers is their apparently contradictory perceptions of the field. On the one hand, some emerging STEM education researchers perceive interactions with the broader DBER as encouraging;  on the other hand, some have the exact opposite experience. This fuels their sense of low-self-efficacy and low sense of belonging. This dichotomy is an interesting reflection of how the community is perceived by newcomers because it suggests that it is not unified in how it defines what it takes to do DBER. To address this, the field  can more explicitly articulate its values, an initiative the STEM DBER alliance introduced  few years ago. This includes working across individual STEM disciplines to present a unified message to individual disciplinary societies and funding agencies as to  what DBER is and does \cite{henderson_towards_2017}. 

When comparing Joiners and Negotiators, we notice the same dichotomies:. Joiners are encouraged by the ease of finding collaborations within the community and feel the support of their departments and institutions, while Negotiators do not yet know how to figure out the norms of collaboration and how to navigate the boundaries between education research and practitioner expertise within STEM departments. Thus, support could include explicit discussions on the dilemmas they experience and ideas they are working out  in order to help researchers imagine or identify their figured worlds as they see fit within their specific socio-cultural contexts. Having explicit discussions about these challenges during professional development activities is an opportunity to support new scholars in the field of education research. Hence, helping emerging STEM education researchers understand and manage the tensions between DBER and home fields of research can help promote persistence and strengthen ties. 

Looking across the three categories Improvers, Joiners, and Negotiators, we notice that a consequential entity in shaping emerging STEM education researchers is the local STEM department. Joiners engaging in action-oriented DBER in a supportive institution are encouraged to expand their research identity into this area, whereas Improvers and Negotiators are navigating challenges involving the perception of doing education research in a traditional STEM department. Improvers who are doing service work and education research are pursuing this work despite the challenge, whereas Negotiators are struggling to know what to do as they navigate this new uncharted territory, especially how to imagine their new, emergent figured world. 

Characterization of these categories expands the current literature in DBER professional development. The Improver and Joiner categories enriches the literature on SFES \cite{bush_disciplinary_2020} by adding some nuance to the experiences of researchers who study and practice education at different types of institutions. Not all of our participants would fall under the SFES characterization as they are not necessarily hired to do education work in their  departments. However, our participants showcase the varied interest and experiences that bring them to discipline-based education research. The Negotiator category adds to the literature on graduate students that discusses new scholars’ professional identity growth \cite{nadelson2017stem}. Nevertheless, our study shows that the uncertainty and questioning that emerging scholars have can also appear at later career stages during academic research transitions. 

Additionally, our three categories about emerging STEM education researchers' conceptualization of DBER align with studies in higher education that investigate academics’ conceptualization of the role of research \cite{akerlind_academic_2008}. With his study participants, Arkerlind identified how doing academic research can be viewed as a job requirement, a personal achievement, a personal exploration space, and as a benefit for larger society. Although our study is framed in the context of STEM education, our findings mirror the job requirement, exploration space, and benefits that Akerlind found with academic research more broadly.  

Nevertheless, our dataset, which brings to the forefront STEM faculty's perceptions of research during their transition to DBER, is distinct from existing literature. Specifically, our findings highlight nuances in challenges and opportunities that are unique to STEM academics. First, there is a special relationship between disciplinary science and education research. For many faculty, interest in DBER stems  from teaching experience in their specific departments and the priorities and challenges facing their department. Second, DBER’s interdisciplinary nature across STEM fields is a special feature that emerging education researchers grapple with. This highlights challenges academics may face navigating research areas outside their PhD or postdoctoral training when associating themselves with new disciplines.

\subsubsection*{Limitations}

By using figured worlds to understand the experiences of emerging STEM education researchers, we centered our analysis on how people negotiate their DBER engagement at the start. However, figured worlds are constantly evolving as participants grow and engage within this space, a characteristic that was brought to the forefront in the Negotiator category. As such it would be fruitful to explore how these emerging researchers continue to engage with DBER and how their figured worlds will change through their participation. Our study focused on emerging STEM education researchers’ positioning and perception of discipline-based education research. As a result we identified categories of DBER figured worlds, which as a by-product contributes to the literature on the field’s culture \cite{hass2021community}. Nevertheless, other studies would be necessary in order to more holistically understand it. In particular,  the perspectives of established members would add more insight on the perception of practices and norms at play. 

Additionally, in future studies, we could build our understanding of disciplinary differences within DBER culture by including other STEM disciplines and more emerging STEM education researchers. This would enable us to see if we missed any nuances that could have been influenced by participation in this professional development program or nuances originating from the STEM disciplines not represented in our current dataset.

\section*{Conclusion}

In this paper, we explored the following research question: how do emerging STEM education researchers position themselves and perceive the role of discipline-based education research? We used figured worlds paired with phenomenography to identify the spectrum of ways emerging STEM education researchers conceptualize their engagement in DBER: to improve their teaching, to make it their new primary research field, and to negotiate how this new research field will fit with their primary one. 

This characterization captures the nuances of how emerging STEM education researchers project themselves in this new space. The challenges and opportunities they are encountering in their engagement portray a wide variety of perceptions of the field colored by their various motivations and circumstances. 
The nuanced perceptions of emerging education researchers allowed us to identify the type of support structures and resources that would benefit them. To help build capacity in STEM education research, professional development opportunities need to take into account DBER’s close ties with disciplinary science to create a smoother and more inclusive transition for scholars in diverse institutional contexts and career stages.

\section*{Acknowledgments}
This work was funded in part by NSF DUE 2025174 and 2039750. Special thanks to the other members of the research group, the PEER facilitation team, and our interview participants for enabling this research.

\nolinenumbers
\bibliography{references}

\end{document}